\documentclass[conference, 9pt]{IEEEtran}
\IEEEoverridecommandlockouts
\usepackage{cite}
\usepackage{amsmath,amssymb,amsfonts}
\usepackage{multirow}
\usepackage{algorithm}
\usepackage{algorithmicx}
\usepackage{algpseudocode}
\usepackage{graphicx}
\usepackage{textcomp}
\usepackage{xcolor}
\usepackage{tabto}
\usepackage{todonotes}
\usepackage{hhline}
\usepackage{enumitem}
\usepackage{setspace}
\usepackage{caption}
\def\BibTeX{{\rm B\kern-.05em{\sc i\kern-.025em b}\kern-.08em
    T\kern-.1667em\lower.7ex\hbox{E}\kern-.125emX}}
\begin{document}

\title{TrojanZero: Switching Activity-Aware Design of Undetectable Hardware \underline{Trojans} with \underline{Zero} Power and Area Footprint}

\author{\IEEEauthorblockN{Imran Hafeez Abbassi\IEEEauthorrefmark{1}, Faiq Khalid\IEEEauthorrefmark{2}, Semeen Rehman\IEEEauthorrefmark{2}, Awais Mehmood Kamboh\IEEEauthorrefmark{1}, \\
		Axel Jantsch\IEEEauthorrefmark{2}, Siddharth Garg\IEEEauthorrefmark{3} and Muhammad Shafique\IEEEauthorrefmark{2}}
\IEEEauthorblockA{\IEEEauthorrefmark{1}School of Electrical Engineering and Computer Science, National University of Sciences and Technology, Islamabad, Pakistan\\
	Email: \{imran.abbasi, awais.kamboh\}@seecs.edu.pk}
\IEEEauthorblockA{\IEEEauthorrefmark{2} Vienna University of Technology, Vienna, Austria\\	
	Email: \{faiq.khalid, semeen.rehman, axel.jantsch, muhammad.shafique\}@tuwien.ac.at}
\IEEEauthorblockA{\IEEEauthorrefmark{3} New York University, New York City, U.S. \\ Email: sg175@nyu.edu}}


\maketitle

\begin{abstract}
Conventional Hardware Trojan (HT) detection techniques are based on the validation of integrated circuits to determine changes in their functionality, and on non-invasive side-channel analysis to identify the variations in their physical parameters. In particular, almost all the proposed side-channel power-based detection techniques presume that HTs are detectable because they only add gates to the original circuit with a noticeable increase in power consumption. \textit{This paper demonstrates how undetectable HTs can be realized with zero impact on the power and area footprint of the original circuit.} Towards this, we propose a novel concept of \textit{TrojanZero} and a systematic methodology for designing undetectable HTs in the circuits, which conceals their existence by gate-level modifications. The crux is to \textit{salvage} the cost of the HT from the original circuit without being detected using standard testing techniques. Our methodology leverages the knowledge of transition probabilities of the circuit nodes to identify and safely remove expendable gates, and embeds malicious circuitry at the appropriate locations with zero power and area overheads when compared to the original circuit. We synthesize these designs and then embed in multiple ISCAS85 benchmarks using a 65nm technology library, and perform a comprehensive power and area characterization. Our experimental results demonstrate that the proposed TrojanZero designs are undetectable by the state-of-the-art power-based detection methods.
\end{abstract}

\begin{IEEEkeywords}
Hardware Trojans, Power Analysis, Area, Signal Probability, ATPG
\end{IEEEkeywords}

\section{Introduction and Related Work}
\label{section_1}
\begin{figure}[h]
	\centering
	\includegraphics[width=1\linewidth]{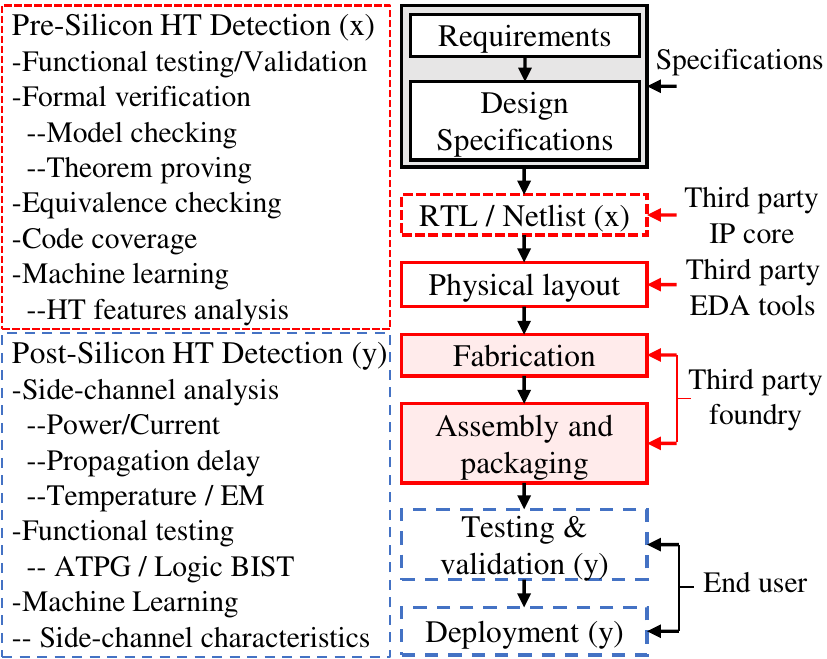} 
	\captionsetup{skip=1pt}
	\caption{IC supply chain stages susceptible to malicious insertions include: (a) Third party IP core (b) EDA tools (c) Untrusted foundries. Pre-silicon detection techniques are employed at the design stage, and post-silicon methods are used at testing and deployment phase.} 
	\label{figure:IC chain}
\end{figure}

The emerging complexity of modern embedded devices and associated cost of advanced CMOS fabrication have increased the trend of outsourcing integrated circuits (ICs) manufacturing processes to untrusted third-parties \cite{xiao2016hardware}. The IC supply chain comprises of various development stages that typically involve untrusted entities such as third-party IP vendors, EDA tools and fabrication foundries as shown in Fig. \ref{figure:IC chain}. Consequently, they are vulnerable to a wide range of HT attacks at some stage of the manufacturing process, which may lead to \textit{leakage} of sensitive information to an adversary, \textit{modification} in functionality, and \textit{degraded performance} of integrated circuits \cite{bhasin2015survey}.
\subsection{HT Detection Techniques}
To mitigate the potential threats of HT attacks in the supply chain, various HT detection techniques have been proposed. Typically, HT detection is performed at the design time (pre-silicon), or after manufacturing (post-silicon) depending on the un-trusted entity involved in the entire process as depicted in Fig. \ref{figure:IC chain}. These techniques are broadly classified into logic-based testing \cite{saha2015improved}, and side-channel analysis \cite{narasimhan2013hardware}.
\paragraph{\textbf{Logic-based Detection}}
Logic-based detection includes equivalence checking \cite{zhang2011case}, and exhaustive simulation \cite{bazzazi2017hardware} which provide 100\% coverage. However, such techniques are not scalable, and applicable only to smaller circuits. Moreover, equivalence checking can only be deployed at the pre-silicon stage. Techniques based on automatic test pattern generation (ATPG) can be used at the post-silicon stage to generate a small set of test vectors that can excite HT trigger nodes \cite{govindan2018logic}. Such techniques maximizes the probability of triggering HTs \cite{chakraborty2009mero}, however, the detection probability and coverage cannot be guaranteed to be 100\%.
\paragraph{\textbf{Side-channel Analysis}}
The detection techniques based on analysis of side-channel measurements are premised on the assumption that HTs distort the parametric profile of the IC. This includes measuring the variations in the observable physical parameters, such as power, delay and temperature to detect any alteration in the IC \cite{di2012side}. Of these, the most commonly used techniques for HT detection are based on power analysis at the post-silicon stage. Some notable works include statistically analyzing the power traces through multiple ports \cite{rad2010sensitivity} to identify HTs. Gate-level characterization using a set of power measurements is proposed in \cite{potkonjak2009hardware} to determine the increase in leakage power due to the addition of malicious gates. Similarly, a sparse gate profiling technique is proposed to detect increase in the leakage power of circuits using statistical learning \cite{chen2017general}.\par
In short, the above-discussed techniques primarily rely on the assumption that HTs are \textit{additive}, i.e., the malicious circuitry is embedded by \textit{adding} gates to the HT-free circuit, resulting in an increase in circuit area and power consumption. \textbf{The goal of this paper is to challenge these underlying assumptions, i.e., are HTs necessarily additive in terms of power and area?}
\begin{figure*}[!t]
	\centering
	\includegraphics[width=\textwidth]{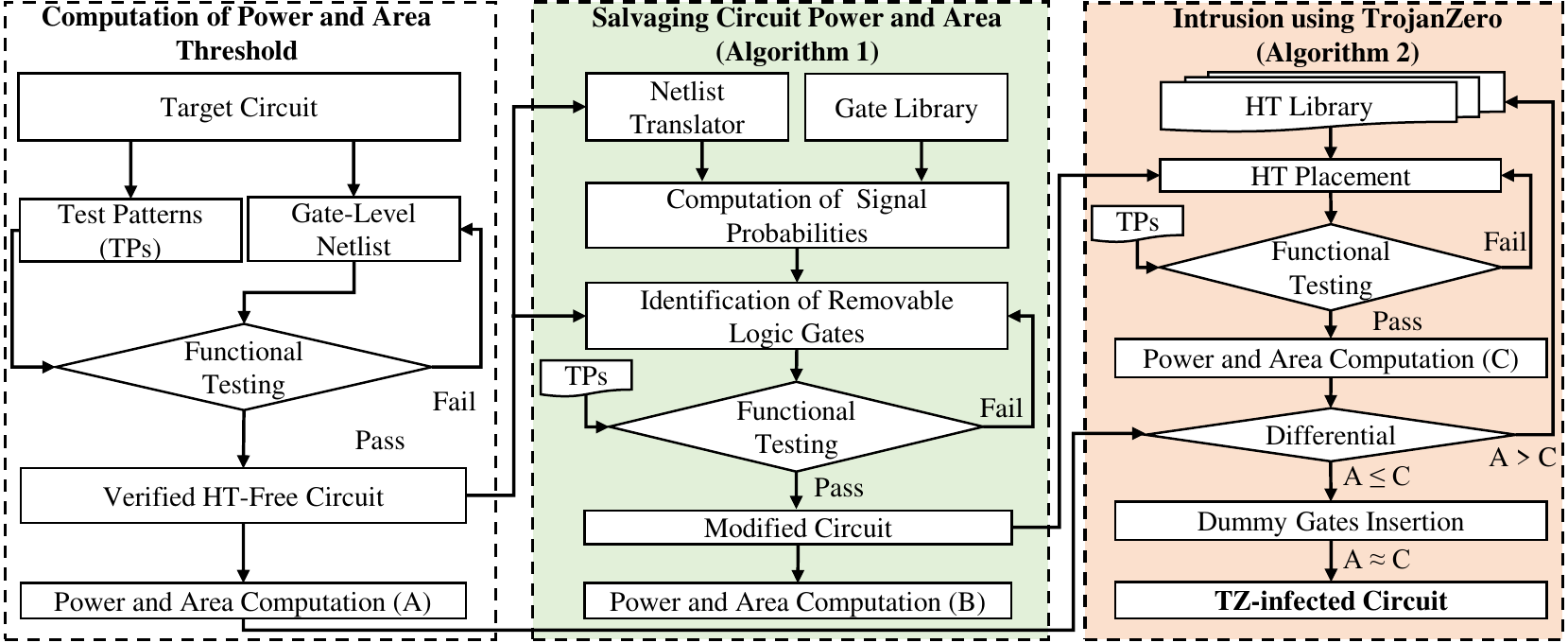} 
	\caption{Proposed flow for our \textit{TrojanZero} Methodology: (a) HT-free circuit is functionally verified using defender's Test Patterns (TPs), and analyzed for computing power and area thresholds. (b) Signal probabilities of the nodes are computed and expendable gates are identified. (c) \textit An HT (i.e., a TrojanZero instance) is embedded in the modified circuit with zero power and area overheads.}
	\label{methodology}
\end{figure*}
\subsection{Motivational Analysis and Research Challenges}
\begin{figure}[H]
	\centering
	\captionsetup{skip=1pt}
	\includegraphics[width=1\linewidth]{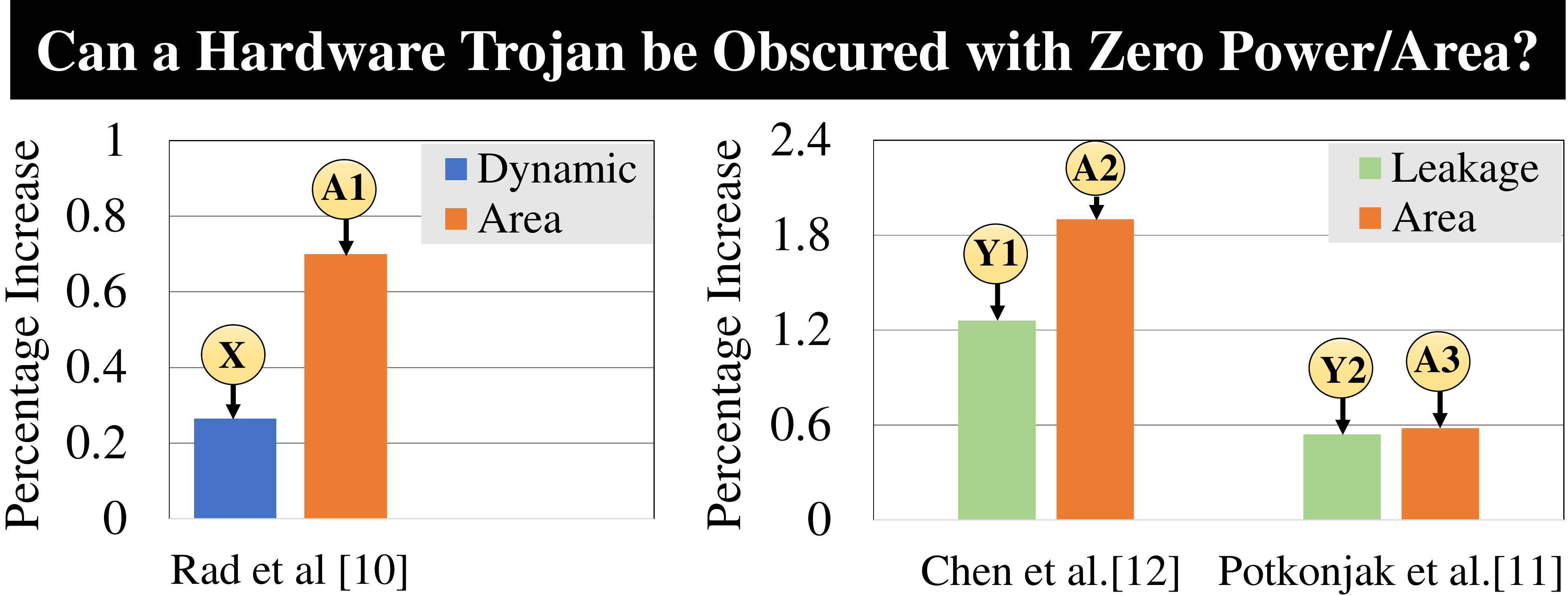} 
	\caption{Minimum power and area overheads that are assumed by state-of-the-art techniques \cite{rad2010sensitivity, chen2017general, potkonjak2009hardware} for successful detection of a single HT in ISCAS benchmark c499.}
	\label{figure:analysis}
\end{figure}

Our analysis in Fig. \ref{figure:analysis} shows the percentage overhead in terms of power and area that is assumed by some of the state-of-the-art methodologies \cite{potkonjak2009hardware,rad2010sensitivity,chen2017general} for successful detection of a single HT in ISCAS benchmark c499. Following key observations are made from this experimental analysis:
\begin{enumerate}[leftmargin=*,wide = 2pt]
	\item Dynamic and leakage power of the HT-infected circuit are perceptibly increased when compared to that of HT-free circuit for successful detection. For instance, the dynamic power of the HT-infected circuit is assumed to exceed at least by 0.265\% as depicted by observation point X. Similarly, the percentage increase assumed for leakage power-based HT detection is shown by Y1 and Y2. 
	\item The area of the circuit is assumed to increase due to presence of an HT. For instance, points A1, A2 and A3 show an increase in the area by 0.7\%, 1.95\% and 0.58\% compared to that of HT-free circuit.
\end{enumerate}
The above-discussed defence techniques appear to work if the increments in power and area are discernible, hence we are going to refrain these additive effects such that HTs are undetected by the existing state-of-the-art methodologies. The open question that is not addressed in the literature is: \textit{how to modify a given circuit in order to insert an HT such that its total power consumption and area are equal to that of a HT-free circuit. We refer to this as devising a \textbf{TrojanZero} methodology with zero power and area overheads.}
\subsection{Attack Model}
In the proposed attack model, we assume that the attacker resides at the foundry where she can modify the circuit in the form of addition, deletion or modification of the gates during fabrication \cite{xiao2016hardware}. Moreover, the attacker acquires the knowledge of specific testing techniques that are used by the defender for the functional validation of circuit after fabrication. The attacker seeks to modify the circuit such that: 
\begin{enumerate}[leftmargin=*,wide = 2pt]
	\item It deviates from the original functionality for certain inputs.
	\item Functionality is not altered on the defender's testing inputs.
	\item Power and area are not increased and HTs are undetectable by the power-based post fabrication analysis.
\end{enumerate}
Prior works have identified two types of HTs: (i) \textit{untargeted} HTs only seek to arbitrarily modify the circuit behavior for certain inputs; and (ii) \textit{targeted} HTs modify the circuit behavior for attacker chosen inputs in an attacker-specified manner. While the primary goal of the \textit{TrojanZero} is to introduce targeted HTs, we will notice that it necessitates the introduction of additional untargeted HTs as well.  
\subsection{Our Novel Contributions}
In this work, we propose a novel concept of \textit{TrojanZero} along with an HT insertion methodology that may subvert the normal operation of ICs, and has \textbf{no additional overheads} in terms of power and area. Our primary contributions in a nutshell are:  
\begin{enumerate}[leftmargin=*,wide = 2pt]
	\item Devising a scheme to identify rarely-activated nodes in the circuit with extremely low signal probabilities.
	\item An algorithm to explore the space of circuit modifications that leave the circuit's functionality on the defender's test patterns unchanged.
	\item A methodology to embed HTs in the target circuit such that there is no increase in total power along with its components, and area. 
	\item Implementation of an HT with a low triggering probability ($<$ 10$^{-4}$) during the functional testing phase. 
\end{enumerate}
\section{Methodology to Design and Implement TrojanZero}
\label{section_2}
The high-level flow of our proposed TrojanZero methodology is shown in Fig. \ref{methodology}. It comprises of three main steps as depicted, and explained in the subsequent sections.
\subsection{Computation of Power and Area Thresholds}
In the first phase, thresholds of the HT-free circuit are computed in terms of power, its components, and area footprint.
\begin{enumerate}[leftmargin=*,wide =2pt]
	\item \textbf{Generating Test Patterns and Functional Verification:}
	A set of defender-specific test patterns is generated for the HT-free circuit through logic-based testing techniques. Functional verification of the circuit is performed by monitoring the outputs against the test patterns (TPs) through simulations. The circuit is re-verified in case of functional failure. 
	\item \textbf{Power and Area Analysis:}
	The verified HT-free circuit is synthesized using the technology library while optimizing it for minimum power. This follows computation of the total power and its constituents, i.e., dynamic and leakage power. Moreover, the area of the circuit is computed in terms of number of gate equivalents (GE) with respect to the technology after synthesis using the ASIC design tools. This analysis of the HT-free circuit is used to specify the power and area thresholds that are to be strictly adhered while embedding HTs. 
\end{enumerate}
\subsection{Salvaging Circuit Power and Area }
We propose Algorithm \ref{algo:1} to salvage power and area from the HT-free circuit with \textit{n} nodes. It executes as follows: \begin{enumerate}[leftmargin=*,wide = 0pt]
	\item \textbf{Inputs:} HT-free circuit ($N$), a set of \textit{q} defender's functional testing algorithms with generated TPs, along with the reference power (P($N$)), and area (A($N$)) computed during analysis. 
	\item \textbf{Computation of Switching Probabilities:}
	The circuit is translated into its corresponding model using a netlist translator, which computes the signal probabilities at each node for being at logic 0 or 1 as depicted in Lines \ref{line2}, and \ref{line3}. For this, we develop a library comprising of basic and complex gates. Each gate computes the probabilities ($P_{g=0}$, $P_{g=1}$) at its output node based on the  probabilities of signals at its inputs. \textit{Similar to other approaches in this field, we also assume that the signal probability at each primary input is 0.5.} 
	\item \textbf{Identification of Expendable Gates:}
	Based on the attacker-specified probability threshold ($P_{th}$), a list of candidate gates ($C$) is obtained that comprises of circuit nodes with signal probabilities close to zero or one as shown in Line \ref{line8}. Each candidate gate is checked for its possible removal from the circuit after detailed functional testing. For this, each node from $C$ with a signal probability close to one is removed and connected to logic 1. Similarly, the output node of gate with signal probability close to zero is connected to logic 0 as depicted in Line \ref{line 14}. After removing a single element from $C$, each of the previous gate is eliminated safely if its output is not connected to any other node of the circuit. This follows detailed functional testing using the defender's $q$ validation algorithms with all TPs. If functional tests are successful then the gates under consideration are successfully removed. However, if any of the test fails, the changes made in the circuit are reverted and next gate from $C$ is tested as depicted in Line \ref{line 23}. This procedure would ensure that such a change in the circuit would go undetected by the defender's post-silicon test techniques. 
	\item \textbf{Power and Area Differential Gains:} After candidates from $C$ are tested and the identified gates are removed, we compute power and area of the modified circuit ($\mathit{N'}$) as shown in Line \ref{line 26}. Moreover, the differential with respect to the HT-free circuit ($\mathit{N}$) is computed to determine the salvaged cost in terms of total power, its components and area.   
\end{enumerate}
\begin{algorithm}[!t]
	\caption{\textbf{:} Salvaging Power and Area}
	\label{algo:1}
	\begin{algorithmic}[1]
		\small
		\Statex \textbf{Input:}
		\Statex $N$ = $\{N_1, N_2, ..., N_n\}$: Verified HT-free circuit with \textit{n} nodes
		\Statex Algo = $\{T_1, T_2, ..., T_q \}$: \textit{q} testing algorithms of defender
		\Statex $P_{th}$ = Attacker specified threshold probability
		\Statex P($N$) = Power of $N$, A($N$) = Area of $N$
		\Statex \textbf{Output:} 
		\Statex $N'$ = $\{N_1', N_2', ..., N_t'\}$: Modified circuit with \textit{t} nodes
		\Statex P($N'$) = Power of $N'$, A($N'$) = Area of $N'$
		\Statex \textbf{Goal:}
		\Statex $\Delta$P = P($N$) - P($N'$), and $\Delta$A = A($N$) - A($N'$) 
		\Statex \textbf{Initialize:}
		i=1, j=1, k=1, m=1; s=1,
		\While{i $\leq$ n} 
		\State $P$($N_i$=0) = $P_{g=0}$, $P_{g=0}$ $\in$ $\{P_{(NAND=0)}$, ..., $P_{(OR=0)}$$\}$; \label{line2}
		\State $P$($N_i$=1) = $P_{g=1}$, $P_{g=1}$ $\in$ $\{P_{(NAND=1)}$, ..., $P_{(OR=1)}$$\}$; \label{line3}
		\If {$P$($N_i$=0) $\geq$ $P_{th}$} X = $\{$$C_1$, $C_2$, ..., $C_j$$\}$; X $\subset$ $N$ 
		\State j=j+1;
		\EndIf
		\If {$P$($N_i$=1) $\geq$ $P_{th}$} Y = $\{$$C_1$, $C_2$, ..., $C_k$$\}$; Y $\subset$ $N$ 
		\State k=k+1;
		\EndIf
		\State $C$ = X $\cup$ Y; i = i + 1;\label{line8} \Comment{Candidate nodes ($C$)}
		\EndWhile
		\While{m$\leq$ j+k} \Comment{Testing each \textit{C(m)} for removal}
		\If {$C$(m) $\in$ X} $C$(m)=0; \Comment{Replace node with 0}\label{line 14} 
		\State Remove preceding gates and update circuit to $N'$;
		\ElsIf
		{$C$(m) $\in$ Y} $C$(m)=1; \Comment{Replace node with 1}
		\State Remove preceding gates and update circuit to $N'$;
		\While{s $\leq$ q}   \Comment{Testing Algorithm 1 to \textit{q}}
		\If {T(s) == Pass} s=s+1;  \Comment{$\forall$ TPs}
		\Else
		\State Revert changes in circuit $N'$; s=q; \label{line 23}
		\EndIf
		\EndWhile
		\EndIf 
		\State m=m+1;
		\EndWhile
		\State Compute P($N'$), A($N'$), $\Delta$P, and $\Delta$A;\label{line 26}
	\end{algorithmic}
\end{algorithm}
\begin{algorithm}[!t]
	\caption{\textbf{:} HT insertion using TrojanZero methodology}
	\label{algo2}
	\begin{algorithmic}[1]
		\small
		\Statex \textbf{Input:}  
		\Statex $N' = \{N_1', N_2', ..., N_t'\}$: Modified circuit with \textit{t} nodes
		\Statex Algo = $\{T_1, T_2, ..., T_q \}$: Set of \textit{q} functional testing algorithms
		\Statex HT $\in$ $\{\{HT_1, HT_2, ..., HT_n\}$: Library with \textit{n} HTs
		\Statex l $\in$ $\{\{l_1, l_2, ..., l_m\}$: \textit{m} potential location in N'
		\Statex P($N$) = Power of HT-free circuit $N$
		\Statex A($N$) = Area of HT-free circuit $N$
		\Statex P($N'$) = Power of  modified circuit $N'$
		\Statex A($N'$) = Area of modified circuit $N'$
		\Statex \textbf{Output:} 
		\Statex $N''$ = TZ-infected circuit
		\Statex P($N''$) = Power of $N''$, A($N''$) = Area of $N''$
		\Statex \textbf{Goal:} 
		\Statex $\Delta$P(TZ) = P($N$) - (P(HT) + P($N'$)) = 0   
		\Statex $\Delta$A(TZ) = A($N$) - (A(HT) + A($N'$)) = 0
		\Statex \textbf{Initialize:}
		i=1, j=1, s=1; 
		\While{i $\leq$ n} 
		\While{j $\leq$ m} \label{Location}
		\While{s $\leq$ q}
		\State Place HT(i) at location l(j);
		\If {T(s) == Pass}  s=s+1;       \Comment{Test with next Algo}
		\Else \; j=j+1;\label{line 8} \Comment {Place HT at next location} 
		\textbf{goto} \ref{Location}
		\EndIf 
		\EndWhile
		\State j = m; \textbf{goto} \ref{compute};
		\EndWhile
		\State Compute P($N''$) = P(HT) + P($N'$); \label{compute}
		\State Compute A($N''$) = A(HT) + A($N'$);
		\If {$\Delta$P(TZ) = 0 \&\& $\Delta$A(TZ) = 0} \label{line 16} 
		\State HT with zero power and area successfully inserted; i = n;
		\Else
		\State i=i+1;
		\EndIf 
		\EndWhile
	\end{algorithmic}
\end{algorithm}
\subsection{HT Insertion using TrojanZero Methodology}
We propose Algorithm \ref{algo2} to insert an HT with zero power and area (TZ), employing our \textit{TrojanZero} methodology. The algorithm provides a procedure for an attacker to systematically exploit the salvaged cost of the modified circuit. This not only makes TZ extremely hard for the defender to be triggered even with bespoke functional test patterns, but renders its existence undetectable from the power and area based analysis techniques.
\begin{enumerate}[leftmargin=*,wide = 0pt]
	\item \textbf{HT Placement:}
	To subvert the desired operation of the circuit, HT from the library of \textit{n} existing malicious circuits is selected and carefully inserted within $N'$. The payload of the HT can be triggered by the attacker-chosen set of vectors provided either by internal or external means. After placement of an HT with an imperceptible trigger for which \textit{m} locations are available for insertion, functional testing is performed with \textit{q} algorithms of the defender. If the test fails, the HT is placed at the next target location as shown in line \ref{line 8}. 
	\item \textbf{Power and Area Analysis:}
	After successful functional testing on defender algorithms, power (P($N''$)) and area (A($N''$)) of the TZ-infected circuit ($\mathit{N''}$) are analyzed to ascertain that it does not surpass the defined thresholds. Conversely, if this is not \textit{true}, then the entire process of HT insertion is repeated by selecting another HT from the library. This follows insertion of the dummy logic gates (if required) to meet the baseline conditions for successful insertion of TZ. These conditions assert that power consumption and area are equivalent to the prescribed thresholds, i.e., $\Delta$P(TZ) = 0, and $\Delta$A(TZ) = 0, as depicted in Line \ref{line 16}. It is mandatory to analyze individual components of power, i.e., dynamic and leakage, independently. These components vary depending upon the location and configuration of HT gates within circuit. \textit {It is plausible that one of the components surpasses the defined threshold, while total power consumption remains within the specified constraints.}            
\end{enumerate}
The insertion of TZ in the modified circuit using Algorithm \ref{algo2} is undetectable to the post-fabrication tests performed by the defender. Therefore, it introduces, (i) targeted \textit{explicit} behavior to modify functionality on attacker specified vectors \cite{haider2017advancing}; (ii) targeted \textit{implicit} behavior while evading side-channel analysis, i.e., power and area. 
\section{Case Study: Intruding 8-bit ALU using TrojanZero}
\label{section_3}
\begin{figure}[!t]
	\centering
	\includegraphics[width=0.4\textwidth]{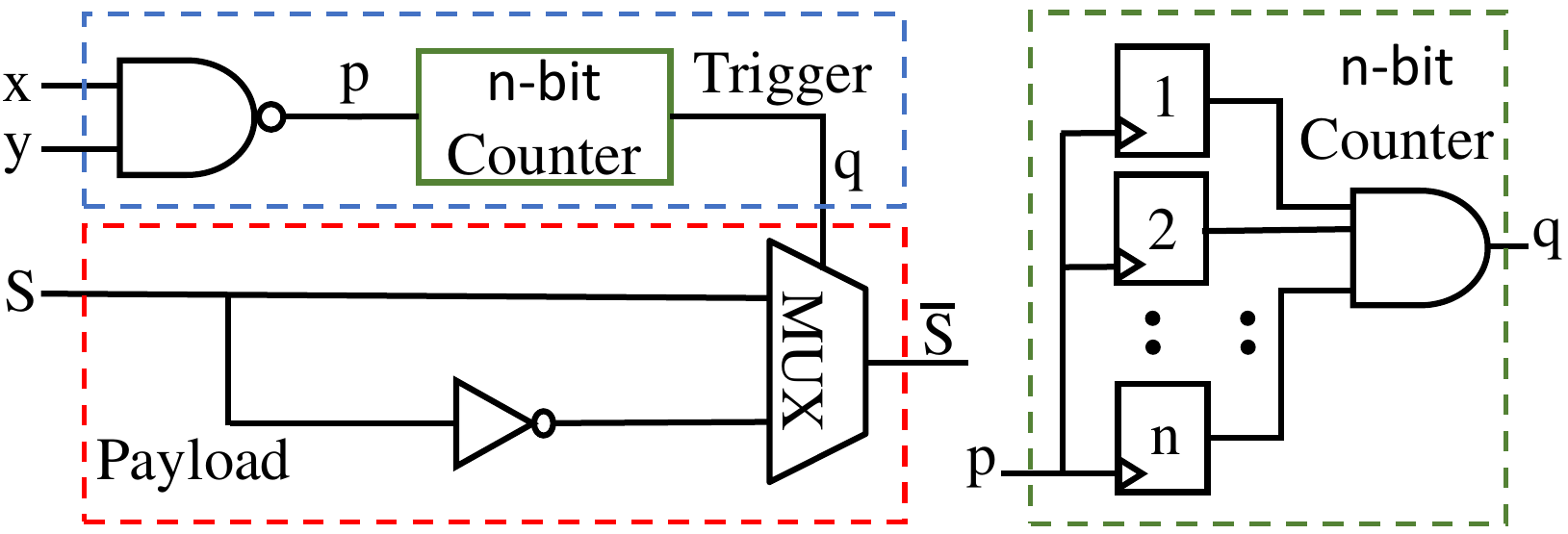} 
	\caption{Asynchronous Counter-based HT \cite{liu2011design}}
	\label{figure:HT}
\end{figure}
\begin{figure*}[!t]
	\centering
	\includegraphics[width=0.9\textwidth]{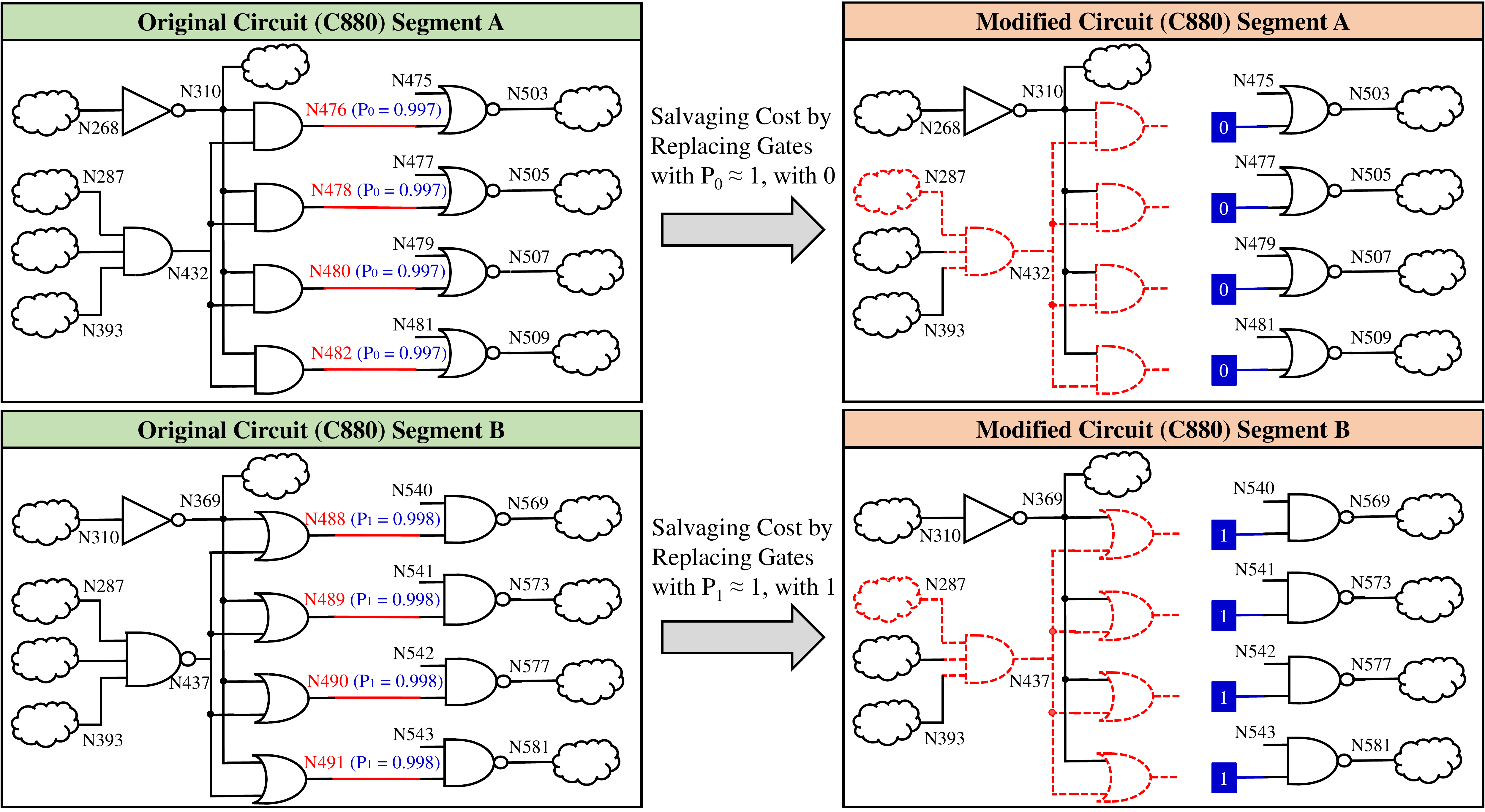} 
	\caption{Expending candidate gates to salvage cost for power and area in ISCAS c880 8-bit ALU.}
	\label{figure:casestudy}
\end{figure*}
To demonstrate the practicability of the proposed approach, we apply the concepts of \textit{TrojanZero} on ISCAS benchmark c880 (8-bit ALU)\cite{iscasc880}.
\subsection{Computation of Power and Area}
We assume that the defender performs functional testing using ATPG stuck-at model. This model is prominently used for diagnostic test generation of transition faults after fabrication, where the nets of circuits are assumed to be stuck at a fixed logic value \cite{bushnell2004essentials}. TPs generated from the test algorithm are used to ascertain the functionality of c880. The total power consumption and area footprint in terms of gate equivalents (GE) is determined as 77.2 $\mu$W, and 365.4 GE, respectively. Similarly, the cell level dynamic and leakage power components are 70.35 $\mu$W, and 6.87 $\mu$W, respectively.
\subsection{Maximizing the Differential of Power and Area} 
We apply Algorithm \ref{algo:1} to salvage the cost in terms of power and area by identifying expendable gates. For each gate of c880, we compute signal probabilities at its output node. We compute the set of candidate gates ($C$) by specifying $P_{th}$= 0.992. Choosing high value of $P_{th}$ provides less number of candidates, however, it increases the ratio of the gates that can be removed from the identified candidates. Moreover, the probability of detecting modifications on defender's bespoke vectors decreases with the higher value of selected $P_{th}$. The set $C$ comprises 27 gate whose signal probabilities are above $P_{th}$ or below $1 - P_{th}$. Fig. \ref{figure:casestudy} shows two segments of c880 comprising of nodes in set $C$ . The set $C$ includes the four AND gates in segment A, i.e., N476, N478, N480 and N482, and the four OR gates highlighted in segment B. Now, we first remove AND (N476) and connect the corresponding input of NOR (N503) to logic 0. This is followed by checking whether any additional independent gates can be removed. However, there is no independent removable gate as AND (N432) is interconnected with adjacent AND gates, and NOT (N310) is connected with other nodes. Next, functional testing is performed and Algorithm \ref{algo:1} determines that this node can indeed be removed from set $C$. Similarly, all other candidate gates are removed iteratively. If, at any step, functional testing fails, the modified circuit is reverted to the previous step. After iterating over all gates in $C$, we test that AND (432) can be expended safely, since all the gates driven by this node have already been removed. All such preceding gates are checked for their secure removal. Similarly, the expendable gates in segment B of c880 are highlighted in Fig. \ref{figure:casestudy}. It is observed from the application of Algorithm \ref{algo:1} on segments A and B that the gate driving node N287 can also be expended. After detailed analysis, we successfully salvaged the cost of 11 logic gates from c880. We compute the power and area of the modified circuit to be 70.2 $\mu$W, and 329.7 GE. This gives us a differential of 7 $\mu$W in power, and 35.7 GE in area footprint that we can used to embed HT.
\subsection{TrojanZero Implementation}
We execute Algorithm \ref{algo2} on the modified circuit to insert an asynchronous counter-based HT \cite{liu2011design} as shown in Fig. \ref{figure:HT}. This HT modifies the signal \textit{S}, whenever the select input \textit{q} of the multiplexer is set to logic 1 by the counter. We placed this HT to modify carry-in (N261) of the c880 ALU on a trigger signal from the counter. The inputs to generate the trigger are provided from rarely-activated nodes of the circuit such that it is not activated during the defender's functional testing. With the insertion of 3-bit counter for trigger generation, it is observed that the total power consumed by the TZ-infected circuit is 76.4 $\mu$W. Moreover, dynamic and leakage power components are 69.32 $\mu$W. and 6.85 $\mu$W, respectively. Similarly, the cell area of the TZ-infected c880 has a footprint of 362.8 GE. Therefore, the outputs of Algorithm \ref{algo2} in terms of \textit{TrojanZero} parameters are: $\Delta$P$_{T}$(TZ) = 0.8 $\mu$W, $\Delta$A(TZ) = 2.6 GE, $\Delta$P$_{D}$(TZ) = 1.03 $\mu$W, $\Delta$P$_{L}$(TZ)= 0.02$\mu$W, where T, D, and L represents total, dynamic and leakage power. The outputs show that TZ-infected circuit has almost equal power and area compared to the HT-free circuit. Therefore, this counter-based HT will not be detected with the state-of-the-art power analysis based HT detection.   
\begin{figure}[!t]
	\centering
	\includegraphics[width=0.47\textwidth]{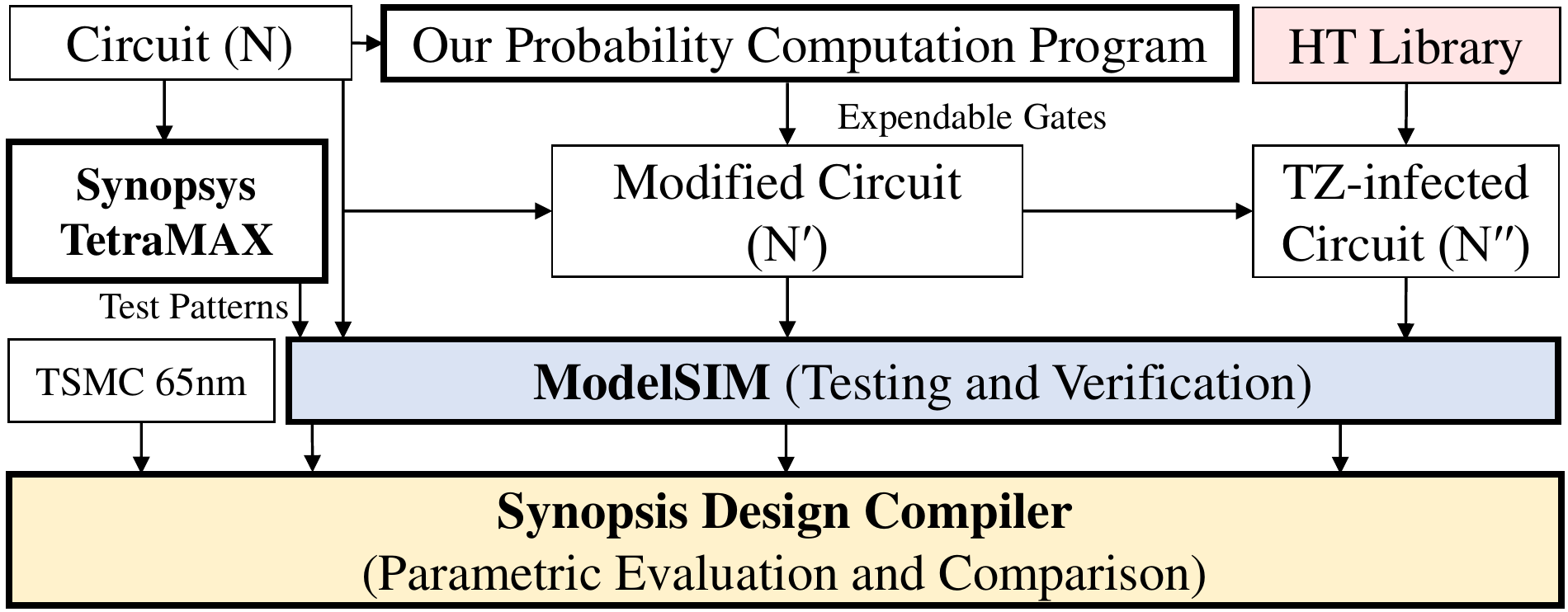} 
	\caption{Tool flow to compute thresholds, salvaging cost by decimating components, and HT insertion.}
	\label{figure:Setup}
\end{figure}
\section{Results and Discussions}
\label{section_4}
We evaluate the proposed methodology on a set of ISCAS85 benchmarks. The experimental setup used for the design and implementation of \textit{TrojanZero} is depicted in Fig. \ref{figure:Setup}. All the simulations are executed on the Red Hat Enterprise Linux (RHEL) 6.8 based computing machine with 8-Core processor $@$ 2.4 GHz, and 16 GB memory, and using the following tools:
\begin{enumerate}[leftmargin=*,wide = 2pt]
	\item \textit{Synopsys TetraMAX (2016)} for automated test generation for the circuits. The circuit is given as input to the tool, which generate the test patterns using the stuck-at model.   
	\item \textit{Matlab 9.1 (R2016b)} based program to compute node probabilities of the HT-free circuit. 
	\item \textit{Mentor Graphics ModelSim 10.5a} for circuit verification during each phase of \textit{TrojanZero} implementation.
	\item \textit{Synopsys Design Compiler (2016)} for synthesizing each circuit using 65nm TSMC technology to perform the detailed analysis based on power and area.
\end{enumerate} 
Table \ref{Results} depicts the $P_{th}$, number of elements in the set $C$, expendable gates ($E_g$), and the estimated probability ($P_{ft}$) for activating HT using random functional testing. The summary of power and area analysis of the HT-free ($\mathit{N}$), modified ($\mathit{N'}$), and TZ-infected ($\mathit{N''}$) circuits is given for comparison. Based on our experimental results, following observations are made:
\begin{table*}[!t]
	\centering
	\renewcommand{\arraystretch}{1}
	\small
	\caption{TrojanZero Analysis for ISCAS85 Benchmarks (I/P = Inputs, $P_{th}$ = Threshold Probability, $C$ = Candidate Gates, $E_g$ = Expendable Gates, $\mathit{N}$ = HT-free circuit, $\mathit{N'}$ = Modified circuit, $\mathit{N''}$ = TZ-infected circuit)}
	\label{Results}
	\begin{tabular}{|c|c|c|c|c|c|c|c|l|c|c|c|c|c|}
		\hline
		\multirow{2}{*}{\textbf{Circuit}} & \multirow{2}{*}{\textbf{Gates}} & \multirow{2}{*}{\textbf{I/P}} & \multirow{2}{*}{\textbf{\begin{tabular}[c]{@{}c@{}}$P_{th}$\end{tabular}}} & \multirow{2}{*}{\textbf{\begin{tabular}[c]{@{}c@{}}$C$\end{tabular}}} & \multirow{2}{*}{\textbf{\begin{tabular}[c]{@{}c@{}}$E_g$\end{tabular}}} & \multirow{2}{*}{\textbf{\begin{tabular}[c]{@{}c@{}} HT\\ (Counter)\end{tabular}}} & \multicolumn{3}{c|}{\textbf{Total Power ($\mu$W)}} & \multicolumn{3}{c|}{\textbf{Area }} & \multirow{2}{*}{\textbf{\begin{tabular}[c]{@{}c@{}}$P_{ft}$\end{tabular}}} \\ \cline{8-13}
		&  &  &  &  &  &  & \textbf{$\mathit{N}$} & \multicolumn{1}{c|}{\textbf{$\mathit{N'}$}} & \textbf{$\mathit{N''}$} & \textbf{$\mathit{N}$} & \textbf{$\mathit{N'}$} & \textbf{$\mathit{N''}$} &  \\ \hline
		c432 & 160 & 32 & 0.975 & 8 & 5 & 2-Bit & 35.6 & 20.83 & 27.7 & 186.8 & 136 & 163 & 0.9 $\cdot$ 10$^{-4}$ \\ \hline
		c499 & 202 & 41 & 0.993 & 12 & 7 & 3-Bit  & 181.9 &173.4 & 177.4 & 463.4 & 396.4 & 451.5 &6.1 $\cdot$ 10$^{-6}$  \\ \hline
		c880 & 383 & 60 & 0.992 & 27 & 11 & 3-Bit & 77.2 & 70.2 & 76.4 & 365.4 & 329.7 & 362.8 & 8.0 $\cdot$ 10$^{-6}$ \\ \hline
		c1908 & 880 & 33 & 0.9986 & 43 & 45 & 5-Bit  & 160.9 & 151.6 & 157.4 & 454.7 & 446.4 & 453.6 & 6.1 $\cdot$ 10$^{-8}$  \\ \hline
		c3540 & 1669 & 50 & 0.992 & 41 & 57 & 5-Bit & 248.5 &  187.2&241.7  & 986.8 & 944.3 & 980 &2.0 $\cdot$ 10$^{-6}$  \\ \hline
	\end{tabular}
\end{table*}
\begin{figure*}[!t]
	\centering
	\includegraphics[width=\textwidth]{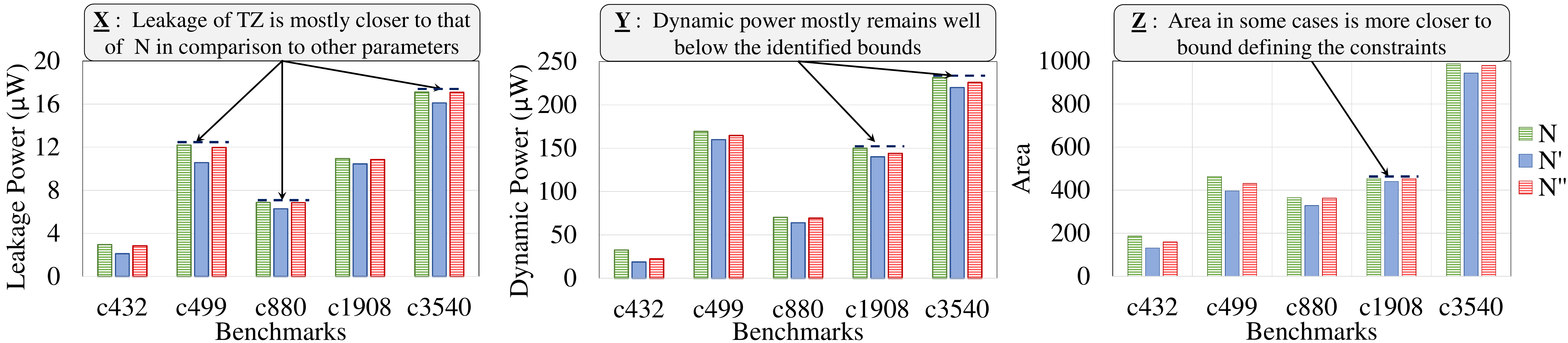} 
	\caption{Comparison of HT-free ($\mathit{N}$), modified ($\mathit{N'}$) and TZ-infected ($\mathit{N''}$) benchmark circuits in terms of area, leakage and dynamic power.}
	\label{results}
\end{figure*}
\begin{enumerate}[leftmargin=*,wide = 0pt]
	\item \textbf{Dynamic, and Leakage Power Analysis:} The power consumption of the circuit along with its components vary subject to the size and location of the inserted HT. This is due to the variability in transition probabilities, and distinct configuration of the different parts of circuit. Fig. \ref{results} presents the comparison of $\mathit{N}$, $\mathit{N'}$ and $\mathit{N''}$ in terms of leakage and dynamic power consumption. It is observed that the leakage power is more liable to violate the boundary of defined constraints compared to the other parameters as depicted by X. Therefore, size of the inserted HT is mainly dictated by its leakage power. With the modifications in the circuit and subsequent insertion of an HT, there is a likelihood that the dynamic power decreases, the leakage power increases, while the total power remains within the defined threshold. This observation stems from the fact that the HT gates leak static power, even when the HT is not triggered. However, the dynamic power consumption of the $\mathit{N''}$ is typically below the defined constraint as depicted by Y in Fig. \ref{results}. Therefore, leakage of the circuit is required to be precisely monitored in all phases.
	\item \textbf{Circuit Configuration and Salvaging Cost:} 
	The admissible size of the HT is proportional to the configuration, complexity, and the salvaged cost of the circuit. There is a potential of inserting multiple HTs of variable sizes in large complex circuit. Typically, complex circuits are likely to have more expendable gates. This is due to the presence of comparatively large number of such nodes that have extremely low probability of transitioning. It is shown in the Table \ref{Results} that the benchmark c1908 comprises of almost half the number of gates compared to that for the c3540, but have 45 expendable gates compared to 57. The configuration of circuit allows for the selection of higher $P_{th}$ and correspondingly more cost is salvaged.
	\item \textbf{Area of TZ-infected Circuit:}
	The observation Z in Fig. \ref{results} shows that there may be a case, where an area cap is required to be adhered more strictly compared to other parameters. Table \ref{Results} shows that at the cost of 45 expendable gates, insertion of 5-bit counter HT in c1908 will have a the margin of 0.2\% from the area cap. However, the same HT has relatively higher margins for dynamic (4.75\%), and leakage (0.8\%) power. Therefore, increasing the size of HT will first violate the area constraint instead of leakage power.
	\item \textbf{Rare-states Combination for HT Trigger:} Choosing the nets with low transition probabilities gives a substantial resistance for triggering an HT on defender's random test vectors as depicted by $P_{ft}$. However, an HT can be triggered on attacker-chosen vectors.
\end{enumerate} \par 
The baseline condition for the successful implementation of the proposed attack is to ensure that power, its components and area $\approx$ 0. In some cases, HT-insertion in the modified circuit may result into negative differential, i.e., discernible decrease of power and area compared to the HT-free circuit. In such cases, dummy gates maybe inserted in parallel to the primary inputs with their outputs unconnected, and thus acquiring negligible differential for all parameters.\par
The TrojanZero methodology relies on the condition that attacker acquires a substantial knowledge pertaining to functional testing techniques of the defender. This scenario is conceivable, since the increasing complexity of  system-on-a-chip (SoC) integration has raised the tendency of outsourcing IC testing services to the third-party vendors. This provides an attacker with a realistic opportunity to obtain relevant information from the third-party. Moreover, the design-for-testability (DfT) techniques e.g., scan-based testing structures provide a reasonable insight to the attacker residing at the foundry about the testing structures employed by the end-user \cite{yasin2017testing}.\par
Apart from the conventional functional testing techniques, the defender may use a set of random (bespoke) vectors for validation which are not known to the attacker. The probability of triggering the targeted HTs using these vectors is very extremely low, as shown by $P_{ft}$ in Table \ref{Results}. Moreover, the probability to reveal un-targeted HTs by the random vectors is determined as follows:  
\begin{equation}
\mathit{P_{u} = \frac{N_u}{2^n}}  
\label{equation:1}
\end{equation}
where, $P_u$ is the probability to trigger an un-targeted HT, $N_u$ represents the number of random input combinations that triggers the untargeted HT, and $n$ is the total number of circuit inputs.\par 
The experimental results and discussion advocate our claims that devising an HT using \textit{TrojanZero} is a pragmatic approach, which can reasonably circumvent the existing state-of-the-art power-based HT detection techniques, and thereby requiring the investigation for new detection methodologies.   
\section{Conclusion}\label{conclusion}
We proposed a novel concept of \textit{TrojanZero} to design and embed undetectable HTs in the target circuit with absolutely no additional costs in terms of power and area. Our method leverages the knowledge of circuit configuration to secure meaningful cost in terms of power and area by decimating its redundant components. We used the salvaged resources to embed HTs in the circuit while adhering with the power, and area cap provided by the analysis of a given HT-free circuit. Our experimental results show that \textit{TrojanZero} can successfully evade the available state-of-the-art HT detection techniques which have the baseline premise that HT insertion eventuates into notable increase of power and size of the circuit. Our methodology provides a foundation for devising new stealthier attacks. An attacker with a reasonable knowledge of circuit configuration can circumvent its security with potentially no risk of getting detected. This instigates a need of exploring more sophisticated and viable techniques for the post-silicon detection of HTs.

\bibliographystyle{ieeetran}
\bibliography{bibliography}

\end{document}